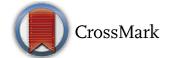

# Improved design of an active landing gear for a passenger aircraft using multi-objective optimization technique

Milad Zarchi[1] · Behrooz Attaran[2]

## Abstract
One of the major subsystems of each airplane is the landing gear system which must be capable of tolerating extreme forces applied to structure during ground maneuver to improve vibration absorbing performance. The traditional landing gear system performs this function well under normal condition, whereas with varying condition of landing and situation of the runway for the airplane, performance of this system decreases noticeably. In this research, for overcoming this problem, the coefficients of controller, the parameters of hydraulic nonlinear actuator added to the traditional shock absorber system, and the vibration absorber are optimized simultaneously by the bee intelligent multi-objective algorithm. As well as, for proving adaptability of this algorithm, this paper presents the sensitivity analysis of three point landing due to the additional payload and the touchdown speed and the robustness analysis of one and two point landings due to the wind conditions as emergency situation on the runway as an innovated work. In order to evaluate the effectiveness and the efficiency of proposed method, the flight dynamic differential equations of an Airbus 320–200 vibrational model during the landing phase are derived and through the numeric technique are solved. The results of numerical analysis for this large-scale airplane model with six degrees of freedom demonstrate that the active shock absorber system in accordance with two types of the bee multi-objective algorithm has good performance in comparison with the passive approach to minimize the bounce displacement and momentum, the pitch displacement and momentum, the suspension travel and impact force in time-domain and frequency-domain by using signal processing, that results in improvement of passenger ride comfort importantly. As well as, enhancement of structure's fatigue life is a likely case as a consequence of study applicable to the industry.

## Nomenclature

| | |
|---|---|
| $e_n(t), e_l(t), e_r(t)$ | Error signals based on PID control logic for the nose, left, and right gears |
| $k_a, k_b$ | Hydraulic coefficients of active control unit |
| $q(t), C_d, w, l, \rho, p_{sh}, p_{sl}$ | Flow fluid flow quantity from servo valve, coefficient of discharge, gradient area of servo valve, displacement of servo valve, density of hydraulic fluid, high pressure in accumulator, low pressure in reservoir |
| $M$ | Sprung mass of the aircraft body |
| $I_{xx}$ | Mass inertia moment about pitch axis |
| $I_{yy}$ | Mass inertia moment about roll axis |
| $m_1$ | Un-sprung mass of nose landing gear |
| $m_2$ | Un-sprung mass of rear left landing gear |
| $m_3$ | Un-sprung mass of rear right landing gear |
| $ks_1, ks_2, ks_3$ | Nose gear sprung mass stiffness rate, rear left gear sprung mass stiffness rate, rear right gear sprung mass stiffness rate |

Springer

| | |
|---|---|
| $cs_1, cs_2, cs_3$ | Nose gear sprung mass damper rate, Rear left gear sprung mass damper rate, rear right gear sprung mass damper rate |
| $kt_1, kt_2, kt_3$ | Nose gear un-sprung mass stiffness rate, rear left gear un-sprung mass stiffness rate, rear right gear un-sprung mass stiffness rate |
| $f_1, f_2, f_3$ | Active control forces based on PID control logic for the nose, left, and right gears |
| $k_p, k_i, k_d$ | PID controller coefficients |
| $z, \theta, \varphi, z_1, z_2, z_3$ | Vertical displacement, pitch displacement, roll displacement of sprung mass and vertical displacement of un-sprung masses |
| $\dot{z}, \dot{\theta}, \dot{\varphi}, \dot{z}_1, \dot{z}_2, \dot{z}_3$ | Vertical velocity, pitch velocity, roll velocity of sprung mass and vertical velocity of un-sprung masses |
| $\ddot{z}, \ddot{\theta}, \ddot{\varphi}, \ddot{z}_1, \ddot{z}_2, \ddot{z}_3$ | Vertical acceleration, pitch acceleration, roll acceleration of sprung mass and vertical acceleration of un-sprung masses |
| $rd_f, rd_{rl}, rd_{rr}$ | Relative displacement for nose, left and right gears |
| $rv_f, rv_{rl}, rv_{rr}$ | Relative velocity for nose, left and right gears |
| $ra_f, ra_{rl}, ra_{rr}$ | Relative acceleration for nose, left and right gears |
| $Fs_1, Fs_2, Fs_3$ | Suspension impact force imparted to the fuselage from nose, left and right gears |
| $Ft_1, Ft_2, Ft_3$ | Tyre impact force imparted to the nose, left and right gears from ground |
| $\{ITAE\}_z, \{ITAE\}_\theta, \{ITAE\}_\varphi, \{ITAE\}_{z_1}, \{ITAE\}_{z_2}, \{ITAE\}_{z_3}$ | The objective function for the displacements of body, nose, left and right gears |
| $\{ITAE\}_{\dot{z}}, \{ITAE\}_{\dot{\theta}}, \{ITAE\}_{\dot{\varphi}}, \{ITAE\}_{\dot{z}_1}, \{ITAE\}_{\dot{z}_2}, \{ITAE\}_{\dot{z}_3}$ | The objective function for the velocities of body, nose, left and right gears |
| $\{ITAE\}_{\ddot{z}}, \{ITAE\}_{\ddot{\theta}}, \{ITAE\}_{\ddot{\varphi}}, \{ITAE\}_{\ddot{z}_1}, \{ITAE\}_{\ddot{z}_2}, \{ITAE\}_{\ddot{z}_3}$ | The objective function for the accelerations of body, nose, left and right gears |
| $\{ITAE\}_{Fs_1}, \{ITAE\}_{Fs_2}, \{ITAE\}_{Fs_3}$ | The objective function for the suspension impact force |
| $\{ITAE\}_{Ft_1}, \{ITAE\}_{Ft_2}, \{ITAE\}_{Ft_3}$ | The objective function for the tyre impact force |
| $[ITAE]_{Type1}$ | The multi-objective function type1 for the sum of displacements, velocities and accelerations |
| $[ITAE]_{Type2}$ | The multi-objective function type2 for the sum of suspension and tyre impact forces |
| $D_N, D_M, V_N, V_M$ | Drag aerodynamic forces and vertical forces |
| $L, W, T$ | Lift aerodynamic force, weight of aircraft and engine force |
| $V, \alpha, \beta$ | Initial vertical velocity when landing, pitch and roll angles of aircraft |

# 1 Introduction and background

The literature review presented in this research gives an overview for how design in the aerospace industry has been done in the past along with emerging ideas, trends, and technology to improve on existing practices. It is also important to note there are many aspects that go into detailed landing gear design, which are not all extensively discussed in this research. This is an emerging trend which the aerospace industry is very interested in pursuing. The philosophy behind utilizing multidisciplinary design solutions as opposed to traditional design practices stems from the belief that each aspect has an effect on the other and optimizing for one aspect of the design may not lead to a design that is optimal in another aspect. Therefore, the goal of this research was to realize this trend by optimizing several components considering both dynamic and structural aspects.

One of the most important and critical subsystems in an aircraft is the landing gear, as it enables aircrafts to gain enough velocity while on the ground to achieve takeoff as well as ensure a safe and smooth touchdown when the destination is reached. To designing the landing gear has become highly sophisticated, because includes different engineering disciplines such as structures, weights, kinematics, economics and runway. The interaction between these different disciplines makes the landing gear a complex system (Roskam 1989). The positioning of the landing gear on the aircraft is limited by several requirements. These requirements include takeoff stability, touchdown stability, wing-tip and engine clearance, ground handling and stability during taxiing. The evaluation of all these limits results in a feasible designing space from which the shortest possible landing gear is found. From the resulting landing gear position, loads on the landing gear struts are calculated. The tyres and the wheels are selected and the brakes and the shock absorbers are designed. For landing gears, it is especially important to look at landing loads, since a hard landing then creates the peaks at high frequency in the shock loads. The process of the designing of a landing gear is extensively documented in the books of Conway, Currey, Roskam and Torenbeek. In the complex engineering systems (such as an aircraft), many different disciplines interact together.

The landing gear is a subsystem that is talented of maximum failure in the air transportation. So, this system is one of the main subsystems that in addition to the requirement to accuracy and high quality in designing, it is required to maximum the maintenance and the periodic inspection for the assurance of the aircraft flight health (Mechant 1978).

There are several methods for investigation of the landing gear dynamic. The best of them is usage from experimental techniques. In the second approach, the landing gear is designed with the same performance of an actual landing gear and then, the measurement and the data collection systems are installed on it. This landing gear is installed under the aircraft. After the installation, different modes of the landing phase are examined and the results of test are obtained and analyzed by these measurement systems. This method is the most accurate for the dynamic simulation of the landing gear (Christofer 2013; Howe 2004).

The dynamic simulation according to the mentioned methods is so costly and time-consuming. At the moment, there are various tools such as ADAMS software that with exclusive properties has the ability of the landing dynamic simulation. But, the modeling in this tool requires the accurate data about the landing gear system that often is not in possession of the designing team in the initial steps of the structure designing and the loading. In the analytical and the semi-analytical methods, each of the gears are modeled one or more mass, the spring and the damper system with various surfaces of accuracy and then, the landing process is simulated by the numerical solution of the motion equations. These methods have high speed for the calculation. As well as, the investigation of various parameters in software or implementation of test becomes easier.

The most aircraft utilize the landing gears with the passive performance that are designed by the producers (Currey 1998). Some researchers (Wignot et al. 1971; Bender and Beiber 1971; McGehee and Garden 1979) have demonstrated that the advantages of the utilizing of the control force to the suspension system to restrict the impact forces imparted to the fuselage. The controllable landing gear system for the first time has been introduced by Ross and Edson 1982. The benefits of this active system in decreasing of the touchdown forces and the oscillations caused by the disturbances have been deduced by (Freymann and Johnson 1985; Freymann 1987) using the analysis and the experimentation. The landing gears with the active controls for a range of the airplane speeds and for various runway surfaces have been investigated by (Catt et al. 1993). A mathematical model has been implemented with the nonlinear characteristics for the main landing gear improved with an external hydraulic system (Horta et al. 1999). The A6 Intruder landing gear system has been surveyed based on the analysis and the test method (Daniels 1996). The NASA researchers (Howell et al. 1991) have surveyed the behavior of the nose shock strut using the F-106B airplane. This study has concentrated on the observation and the experimental data based on the drop tests and has deduced that the active landing gear system significantly increases the aircraft's structure fatigue life during the ground maneuvers. A mathematical model of two degrees of freedom for the A6 Intruder has been investigated with the single active landing gear system. It pointed to improvement of the shock absorber performance with the proportional integral derivative controller that the gains have optimized through the numerical simulation experiments (Wang et al. 2008).

According to the studies of Karnopp (1983) for the automotive applications, compared with the passive control, the active control has had significant adaptability. The aircraft vibration due to the runway irregularities has been analyzed with the classical controllers whose the coefficients have been tuned by the Ziegler-Nichols method (Sivakumar and Haran 2015a, b; Toloei et al. 2016). The performance improvement of an active vibration absorber subsystem for a Fokker aircraft model using a bee algorithm based on the multi-objective intelligent optimization during taxiing phase by some searchers (Zarchi and Attaran 2017) has been done.

The application of bee algorithm (BA) in the finite element (FE) model updating of the structures has been investigated (Moradi et al. 2010). An artificial Bee Colony Algorithm (ABCA) has been used to obtain the optimal size and the location of viscous dampers in the planar buildings to reduce the damage to the frame systems during an earthquake (Sonmez et al. 2013). The objective has been to obtain the optimum design for the reinforced concrete continuous beams in terms of the cross section dimensions and the reinforcement details using a fine tuned Artificial Bee Colony (ABC) Algorithm while still satisfying the constraints of the ACI Code. The ABC algorithm used by these researchers (Jahjouh et al. 2013) has been slightly modified to include a Variable Changing Percentage (VCP) that further improves its performance when dealing with the members consisted of the multiple variables. The aim of the study done by Sonmez 2011 has been to present an optimization algorithm based on the ABC algorithm for the discrete optimum designing of the truss structures. Stolpe 2011 has compared the Artificial Bee Colony algorithm numerically to three alternative heuristics on the same benchmark examples by Sonmez 2011.

In the present work, the dynamic equations for the collection of three airplane landing gears are achieved that each of them are modeled with one spring and damper in parallel as the shock absorber and one spring as the tyre. As well as, various scenarios of the landing in different situations are considered. These scenarios are mentioned in the aerial standards as the landing main conditions. In present research, the approximate specifications of Airbus 320–200 aircraft are utilized for achieving of the numerical solution and comparison with the standards. So, the standard requirements according to the Federal Aviation Regulation (FAR) are employed. The landing scenarios for the airplane with the nose and the main landing gears are sorted to five parts as follows:

1. Three point landing
2. Two point landing
3. One point landing
4. Lateral load landing
5. Rebound landing

As development cycles and prototyping iterations begin to decrease in the aerospace industry, it is important to develop and improve practical methodologies to meet all design metrics. This research presents an efficient methodology that applies high-fidelity multi-disciplinary design optimization technique to passenger aircraft landing gear assembly, for improvement of touchdown performance. The design optimization process utilized in this research was sectioned into two main stages: setup of mathematical model and multi-objective optimization algorithm.

In this research, first the landing gear designing literature available is identified in chapter 1. Then in chapter 2, the overall landing gear vibrational modeling process and the motion equations is explained. Chapter 3 describes an overview of the PID controller design by using two methods consist of the Ziegler-Nichols and bee optimization algorithm. Then the current innovation in the field of the optimal suspension system and the optimal industrial controller design and the analysis is presented. Chapter 4 includes the numerical simulation for a special aircraft during the landing phase with three scenarios and chapters 5 and 6 cover conclusion and the future work.

## 2 A vibrational model of the Airbus320–200 with the active suspension subsystem during the landing phase

One of the most important and critical subsystems in an aircraft is the landing gear, as it enables aircrafts to gain enough velocity while on the ground to achieve takeoff as well as ensure a safe and smooth touchdown when the destination is reached. Despite being such a heavily loaded structure, this subsystem typically makes up only 2.5% to 5% of the weight of an aircraft and roughly 1.5% of the overall initial sunk cost. However, it has been determined that approximately 50% of the total damages that incur on a commercial aircraft over its lifetime are related to taxi, takeoffs and landing, which correspond to 20% of the direct maintenance cost (Roskam 1989). Given these facts, one can start to see some of the challenges that designers may face when trying to developing improvements to the existing system. In order to gain a better understanding of these challenges, a basis of the fundamentals of landing gear design must be established, and will be covered in the following subsections.

### 2.1 Description of the model

In order to gain a better understanding of the landing gear design process, the nomenclature typically used to identify each landing gear component and its functionality should be defined. A generic retractable and telescopic landing gear system is shown below in Fig. 1. Note that certain features and components may vary from one aircraft to the other (Roskam 1989).

The active combination consists of a servo valve, a low-pressure (LP) reservoir, a high-pressure (HP) accumulator, a hydraulic pump, an electronic controller and the feedback transducers (Wang et al. 2008; Sivakumar and Haran 2015a, b; Toloei et al. 2016a). When an aircraft lands, the shock absorber stroke is influenced by the aircraft's payload and varies depending on the runway characteristics and the vertical velocity at touchdown. The stroke is measured by the transducers and their signals input into the electronic controller. Figure 2 shows the six degrees of freedom aircraft model. There are three degrees of freedom for roll, pitch and the vertical displacement (bouncing) of the sprung mass or the airframe and three for the vertical displacement of the un-sprung masses or the landing gears. The active control forces are represented by $f_1$, $f_2$, $f_3$. In this model, the yaw degree of freedom is not considered and the airplane has just longitudinal motion.

### 2.2 Mathematical model of the airplane and landing gear with active control unit

From the kinematic considerations, the relative displacement for nose, left and right gears respectively are given by

$$rd_f = z - a\theta - (d-e)\varphi - z_1 \\ rd_{rl} = z + b\theta - d\varphi - z_2 \\ rd_{rr} = z + b\theta + e\varphi - z_3 \quad (1)$$

In a similar manner, the relative velocity for nose, left and right gears respectively are given by

$$rv_f = \dot{z} - a\dot{\theta} - (d-e)\dot{\varphi} - \dot{z}_1 \\ rv_{rl} = \dot{z} + b\dot{\theta} - d\dot{\varphi} - \dot{z}_2 \\ rv_{rr} = \dot{z} + b\dot{\theta} + e\dot{\varphi} - \dot{z}_3 \quad (2)$$

Finally, the relative acceleration for nose, left and right gears respectively are given by

$$ra_f = \ddot{z} - a\ddot{\theta} - (d-e)\ddot{\varphi} - \ddot{z}_1 \\ ra_{rl} = \ddot{z} + b\ddot{\theta} - d\ddot{\varphi} - \ddot{z}_2 \\ ra_{rr} = \ddot{z} + b\ddot{\theta} + e\ddot{\varphi} - \ddot{z}_3 \quad (3)$$

Consideration of balance of the forces and moments yields the equations of motion for the aircraft in touchdown phase for the bouncing, pitching, and rolling motions of the body:

**Fig. 1** Main components of landing gear system (Roskam 1989)

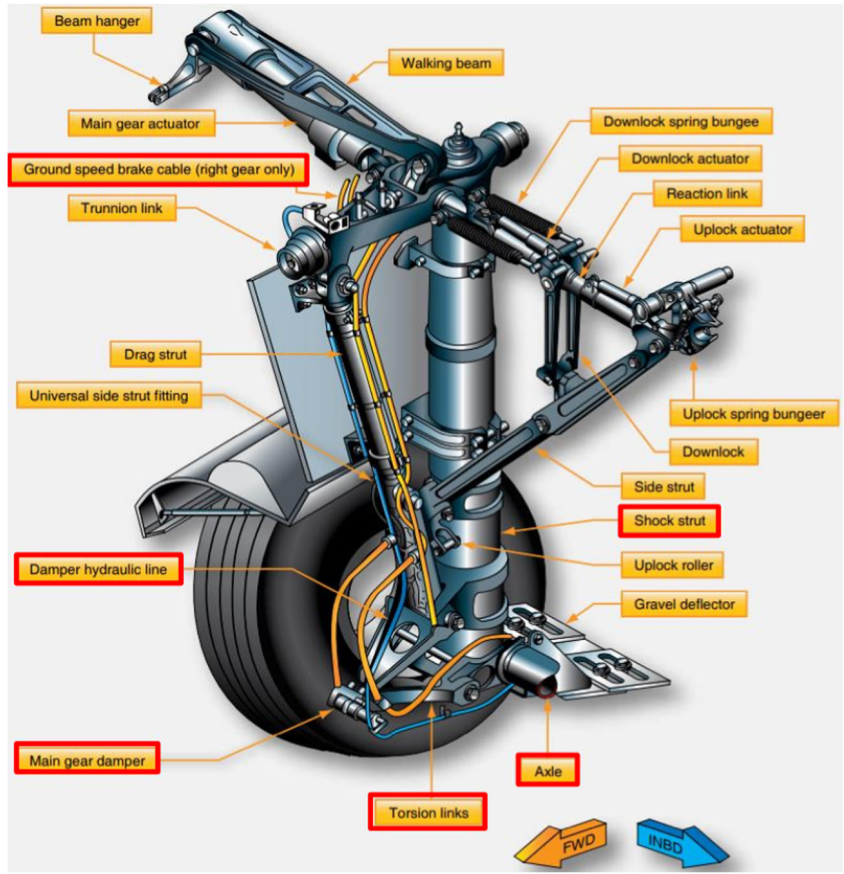

For the bouncing motion of the aircraft body

$$M\ddot{z} + ks_1 rd_f + ks_2 rd_{rl} + ks_3 rd_{rr} + cs_1 rv_f + cs_2 rv_{rl} + cs_3 rv_{rr} + f_1 + f_2 + f_3 = 0 \quad (4)$$

For the pitching motion of the aircraft body

$$I_{xx}\ddot{\theta} - ks_1 rd_f a + ks_2 rd_{rl} b + ks_3 rd_{rr} b - cs_1 rv_f a + cs_2 rv_{rl} b + cs_3 rv_{rr} b - f_1 a + f_2 b + f_3 b = 0 \quad (5)$$

For the rolling motion of the aircraft body

$$I_{yy}\ddot{\varphi} - ks_1 rd_f(d-e) + ks_2 rd_{rl} d + ks_3 rd_{rr} e - cs_1 rv_f(d-e) + cs_2 rv_{rl} d + cs_3 rv_{rr} e - f_1(d-e) - f_2 d + f_3 e = 0 \quad (6)$$

**Fig. 2** A vibrational model of Airbus320–200 airplane (Zarchi and Attaran 2017)

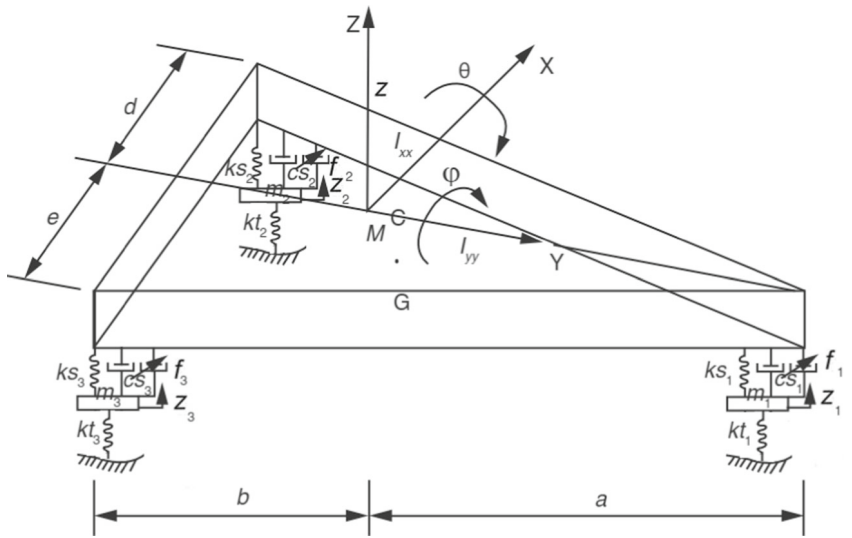

The vertical displacements of the landing gear masses are given as follows.

For the nose gear mass

$$m_1\ddot{z}_1 - k_{s_1}rd_f - c_{s_1}rv_f + k_{t_1}z_1 - f_1 = 0 \quad (7)$$

For the left gear mass

$$m_2\ddot{z}_2 - k_{s_2}rd_{rl} - c_{s_2}rv_{rl} + k_{t_2}z_2 - f_2 = 0 \quad (8)$$

For the right gear mass

$$m_3\ddot{z}_3 - k_{s_3}rd_{rr} - c_{s_3}rv_{rr} + k_{t_3}z_3 - f_3 = 0 \quad (9)$$

Collecting the Eqs. (1–9), a matrix system of equations of the following form is obtained:

$$[m]\{\ddot{z}\} + [c]\{\dot{z}\} + [k]\{z\} = \{f\} \quad (10)$$

The mass matrix is defined by

$$[m] = \begin{bmatrix} M & 0 & 0 & 0 & 0 & 0 \\ 0 & I_{xx} & 0 & 0 & 0 & 0 \\ 0 & 0 & I_{yy} & 0 & 0 & 0 \\ 0 & 0 & 0 & m_1 & 0 & 0 \\ 0 & 0 & 0 & 0 & m_2 & 0 \\ 0 & 0 & 0 & 0 & 0 & m_3 \end{bmatrix} \quad (11)$$

The damping matrix is presented by

$$[c] = \begin{bmatrix} \lambda_1 & \lambda_2 & \lambda_3 & -c_{s_1} & -c_{s_2} & -c_{s_3} \\ \lambda_2 & \lambda_4 & \lambda_5 & ac_{s_1} & -bc_{s_2} & -bc_{s_3} \\ \lambda_3 & \lambda_5 & \lambda_6 & (d-e)c_{s_1} & dc_{s_2} & -ec_{s_3} \\ -c_{s_1} & ac_{s_1} & (d-e)c_{s_1} & c_{s_1} & 0 & 0 \\ -c_{s_2} & -bc_{s_2} & dc_{s_2} & 0 & c_{s_2} & 0 \\ -c_{s_3} & -bc_{s_3} & -ec_{s_3} & 0 & 0 & c_{s_3} \end{bmatrix} \quad (12)$$

Where,

$$\begin{aligned}
\lambda_1 &= cs_{NLG} + cs_{LLG} + cs_{RLG} = cs_1 + cs_2 + cs_3 \\
\lambda_2 &= -acs_{NLG} + bcs_{LLG} + bcs_{RLG} = -acs_1 + bcs_2 + bcs_3 \\
\lambda_3 &= -(d-e)cs_{NLG} - dcs_{LLG} + ecs_{RLG} = -(d-e)cs_1 - dcs_2 + ecs_3 \\
\lambda_4 &= a^2 cs_{NLG} + b^2 cs_{LLG} + b^2 cs_{RLG} = a^2 cs_1 + b^2 cs_2 + b^2 cs_3 \\
\lambda_5 &= (d-e)acs_{NLG} - dbcs_{LLG} + ebcs_{RLG} = (d-e)acs_1 - dbcs_2 + ebcs_3 \\
\lambda_6 &= (d-e)^2 cs_{NLG} + d^2 cs_{LLG} + e^2 cs_{RLG} = (d-e)^2 cs_1 + d^2 cs_2 + e^2 cs_3
\end{aligned} \quad (13)$$

The stiffness matrix is given by

$$[k] = \begin{bmatrix} \mu_1 & \mu_2 & \mu_3 & -k_{s_1} & -k_{s_2} & -k_{s_3} \\ \mu_2 & \mu_4 & \mu_5 & ak_{s_1} & -bk_{s_2} & -bk_{s_3} \\ \mu_3 & \mu_5 & \mu_6 & (d-e)k_{s_1} & dk_{s_2} & -ek_{s_3} \\ -k_{s_1} & ak_{s_1} & (d-e)k_{s_1} & k_{s_1} + k_{t_1} & 0 & 0 \\ -k_{s_2} & -bk_{s_2} & dk_{s_2} & 0 & k_{s_2} + k_{t_2} & 0 \\ -k_{s_3} & -bk_{s_3} & -ek_{s_3} & 0 & 0 & k_{s_3} + k_{t_3} \end{bmatrix} \quad (14)$$

Where,

$$\begin{aligned}
\mu_1 &= ks_{NLG} + ks_{LLG} + ks_{RLG} = ks_1 + ks_2 + ks_3 \\
\mu_2 &= -aks_{NLG} + bks_{LLG} + bks_{RLG} = -aks_1 + bks_2 + bks_3 \\
\mu_3 &= -(d-e)ks_{NLG} - dks_{LLG} + eks_{RLG} = -(d-e)ks_1 - dks_2 + eks_3 \\
\mu_4 &= a^2 ks_{NLG} + b^2 ks_{LLG} + b^2 ks_{RLG} = a^2 ks_1 + b^2 ks_2 + b^2 ks_3 \\
\mu_5 &= (d-e)aks_{NLG} - dbks_{LLG} + ebks_{RLG} = (d-e)aks_1 - dbks_2 + ebks_3 \\
\mu_6 &= (d-e)^2 ks_{NLG} + d^2 ks_{LLG} + e^2 ks_{RLG} = (d-e)^2 ks_1 + d^2 ks_2 + e^2 ks_3
\end{aligned} \quad (15)$$

And the force vector is

$$\{f\} = \begin{Bmatrix} -f_1 - f_2 - f_3 \\ f_1 a - f_2 b - f_3 b \\ f_1 (d-e) + f_2 d - f_3 e \\ f_1 \\ f_2 \\ f_3 \end{Bmatrix} \quad (16)$$

## 3 Design of the proportional-integration-derivative control law

The focus of this section is to perform high-fidelity multidisciplinary design optimization to satisfy structural and dynamic performance requirements. Design optimization is considered to be the process of determining the best set of parameters to yield the greatest performance gain, while satisfying a particular set of rules or constraints. The active control force caused by the suspension system and the control law for the improvement of system performance is defined according to (Toloei et al. 2016b; Zarchi and Attaran 2017).

The active control force and the flow quantity are computed by

$$\begin{aligned} f_{active}(t) &= k_a q(t) + k_b q(t)|q(t)| \\ q(t) &= C_d w l \sqrt{\frac{p_{sh} - p_{sl}}{\rho}} \end{aligned} \quad (17)$$

The corresponding active control forces based on PID control logic for the nose, left, and right gears are given by

$$\begin{aligned}
f_1(t) &= k_p e_n(t) + k_i \int_0^t e_n(t) + k_d \frac{de_n(t)}{dt} \\
f_2(t) &= k_p e_l(t) + k_i \int_0^t e_l(t) + k_d \frac{de_l(t)}{dt} \\
f_3(t) &= k_p e_r(t) + k_i \int_0^t e_r(t) + k_d \frac{de_r(t)}{dt}
\end{aligned} \quad (18)$$

The error signal is given by

$$\begin{aligned}
e_n(t) &= \dot{r}(t) - \left(\dot{z} - a\dot{\theta} - (d-e)\dot{\varphi} - \dot{z}_1\right)(t) \\
e_l(t) &= \dot{r}(t) - \left(\dot{z} + b\dot{\theta} - d\dot{\varphi} - \dot{z}_2\right)(t) \\
e_r(t) &= \dot{r}(t) - \left(\dot{z} + b\dot{\theta} + e\dot{\varphi} - \dot{z}_3\right)(t)
\end{aligned} \quad (19)$$

The output signals at the gears that represent the displacement of the servo valves are given by

$$l_1(t) = k_p\{\dot{r}(t)-(\dot{z}-a\dot{\theta}-(d-e)\dot{\varphi}-\dot{z}_1)(t)\} + k_i\{r(t)-(z-a\theta-(d-e)\varphi-z_1)(t)\}$$
$$+k_d\{\ddot{r}(t)-(\ddot{z}-a\ddot{\theta}-(d-e)\ddot{\varphi}-\ddot{z}_1)(t)\}$$
$$l_2(t) = k_p\{\dot{r}(t)-(\dot{z}+b\dot{\theta}-d\dot{\varphi}-\dot{z}_2)(t)\} + k_i\{r(t)-(z+b\theta-d\varphi-z_2)(t)\}$$
$$+k_d\{\ddot{r}(t)-(\ddot{z}+b\ddot{\theta}-d\ddot{\varphi}-\ddot{z}_2)(t)\}$$
$$l_3(t) = k_p\{\dot{r}(t)-(\dot{z}+b\dot{\theta}+e\dot{\varphi}-\dot{z}_3)(t)\} + k_i\{r(t)-(z+b\theta+e\varphi-z_3)(t)\}$$
$$+k_d\{\ddot{r}(t)-(\ddot{z}+b\ddot{\theta}+e\ddot{\varphi}-\ddot{z}_3)(t)\}$$

(20)

### 3.1 Ziegler-Nichols proportional integral derivative tuning

The values of the parameters $K_p$, $K_i$ and $K_d$ according to Ziegler-Nichols (ZN) tuning rules are introduced in Table 1 (Ziegler and Nichols 1942).

### 3.2 The bee intelligent optimization algorithm

#### 3.2.1 The bee in nature

A colony of the honey bees can be seen as a diffuse creature which can extend itself over long distances in the multiple directions in order to exploit a large number of food sources at the same time. In principle, the flower patches with plentiful amounts of nectar or pollen that can be collected with less effort should be visited by more bees, whereas patches with less nectar or pollen should receive fewer bees.

The foraging process begins in a colony by the scout bees being sent to search for promising the flower patches. The scout bees search randomly from one patch to another. During the harvesting season, a colony continues its exploration, keeping a percentage of the population as the scout bees (Frisch 1976).

When they return to the hive, those scout bees that found a patch which is rated above a certain threshold (measured as a combination of some constituents, such as sugar content) deposit their nectar or pollen and go to the "dance floor" to perform a dance known as the "waggle dance" (Seeley 1996). This dance is essential for the colony communication, and contains three vital pieces of the information regarding a flower patch: the direction in which it will be found, its distance from the hive and its quality rating (or fitness) (Frisch 1976). This information helps the bees to find the flower patches precisely, without using guides or maps. Each individual's knowledge of the outside environment is gleaned solely from the waggle dance. This dance enables the colony to evaluate the relative merit of the different patches according to both the quality of the food they provide and the amount of energy needed to harvest it (Camazine et al. 2003). After the waggle dancing on the dance floor, the dancer bee (i.e., the scout bee) goes back to the flower patch with follower bees that were waiting inside the hive. The number of follower bees assigned to a patch depends on the overall quality of the patch. This allows the colony to gather the food quickly and efficiently.

While harvesting from a patch, the bees monitor its food level. This is necessary to decide upon the next waggle dance when they return to the hive (Camazine et al. 2003). If the patch is still good enough as a food source, then it will be advertised in the waggle dance and more bees will be recruited to that source.

#### 3.2.2 Description of the bee algorithm

As mentioned above, the bee algorithm is an optimization algorithm inspired by the natural foraging behavior of the honey bees (Pham and Ghanbarzadeh 2007). The steps 1 to 8 show the pseudo code of the basic bee algorithm in its simplest form (Pham et al. 2006b–e). The algorithm requires a number of parameters to be set, namely: the number of scout bees (N), the number of patches selected out of N visited points (M), the number of elite patches out of m selected patches (E), the number of bees recruited for the best E patches (Nep), the number of bees recruited for the other (M-E) selected patches (Nsp), the size of patches (Ngh) and stopping criterion (Pham et al. 2007b–g; Salem et al. 2009). The algorithm starts with N scout bees being placed randomly in the search space. The fitness of the points visited by the scout bees are evaluated in the step 2. In the step 3, the bees that have the highest fitness are designated as "selected bees" and the sites visited by them are chosen for the neighborhood search. Then, in the steps 4 and 5, the algorithm conducts searches in the neighborhood of the selected bees, assigning more bees to search near to the best E bees. The bees can be chosen directly according to the fitness associated with the points they are visiting. Alternatively, the fitness values are used to determine the probability of the bees being selected. The searches in the neighborhood of the best E bees which represent more promising solutions are made more detailed by the recruiting more bees to follow them than the other selected bees. Together with scouting, this differential recruitment is a key operation of the bee algorithm.

In the step 6, for the each site only the bee with the highest fitness will be selected to form the next bee population. In nature, there is no such restriction. This constraint is introduced here to reduce the number of points to be explored. In the step 7, the remaining bees in the population are assigned randomly around the search space scouting for the new potential solutions. These steps are repeated until a stopping criterion is met. At the end of the each iteration, the colony will have two parts to its new population representatives from the each selected patch and other scout bees assigned to conduct the random searches. The flowchart of these steps is presented in Fig. 3.

**Table 1** PID controller coefficients according to ZN method

| Type of landing gear | $K_p$ | $K_i$ | $K_d$ |
| --- | --- | --- | --- |
| Nose landing gear | 0.2 | 2.5 | 0.0076 |
| Main landing gear | 0.5 | 275 | 0.0003 |

### 3.2.3 Advantages of the bee algorithm

According to (Moradi et al. 2010) about the application of bee algorithm (BA) in the finite element (FE) model updating of structures, (Toloei et al. 2016a) about the application of bee algorithm (BA) for the adoptable landing gear vibration behavior and to design proportional integration derivative (PID) classical techniques for control of active hydraulic nonlinear actuator, (Toloei et al. 2016b) about the application of bee algorithm (BA) for the model of semi-active suspension system that chooses damping performance of shock absorber at touchdown to be the purpose of fuzzy control on landing gear, (Toloei et al. 2017) about the application of bee algorithm (BA) for the advanced vibrational model for study of the adjustable vibration absorber and to plan proportional-integration-derivative approach for adapting semi-active control force, (Zarchi and Attaran 2017) about the application of bee algorithm (BA) for designing a proportional-integral-derivative technique for control of an active vibration absorber system using a hydraulic nonlinear actuator for the six-degree-of-freedom aircraft model in the taxiing phase, the major advantages which the bee algorithm holds over other optimization algorithms are:

i. Simplicity, flexibility and robustness
ii. Use of fewer control parameters compared with many other search techniques
iii. Ease of hybridization with other optimization algorithms
iv. Ability to handle the objective cost with stochastic nature
v. Ease of implementation with basic mathematical and logical operations
vi. Finding the global optimization solution

### 3.2.4 Implementation of the bee multi-objective algorithm

Here, the bee multi-objective clever algorithm is applied for the tuning of PID industrial controller parameters according to the steps 1 to 12 (Pham et al. 2005, 2006a). In this paper, this algorithm is implemented by using Integral of the Time weighted Absolute value of the Error (ITAE) in accordance with the Eqs. 21 to 29 as the optimization function (Pham et al. 2007a). The convergence plot of two multi-objective functions versus iteration (that is function of time) is displayed in Fig. 5. The variables of this method and PID controller, high and low pressures, damping and stiffness coefficients for the active vibration absorber system are obtained in Tables 2, 3 and 4. These parameters will be used later to compare the behavior of the landing gear during the simulation. The flowchart of bee multi-target algorithm for the tuning of the coefficients according to objective functions type 1 and type 2 is described in Fig. 4.

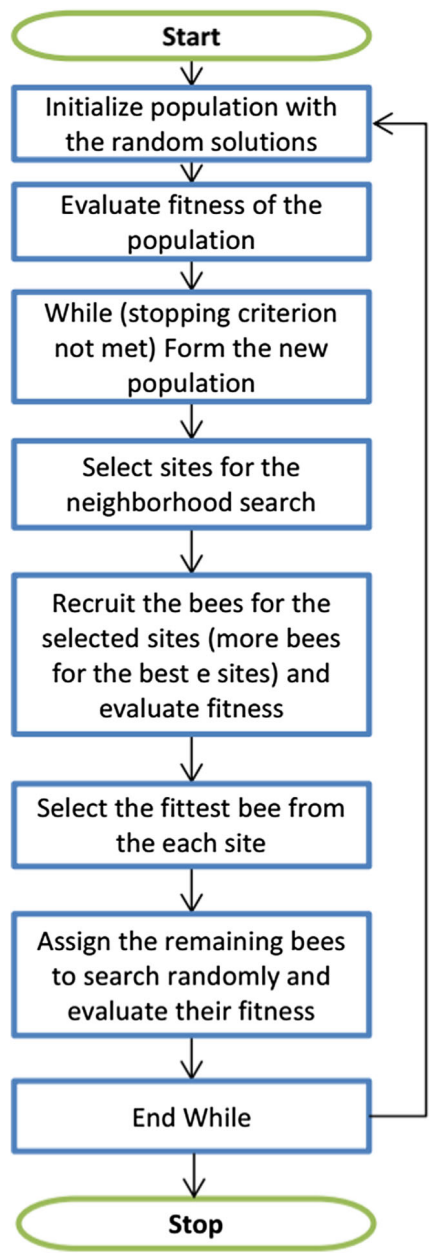

**Fig. 3** Flowchart of bee proposed algorithm

Table 2  The parameters used in BA

| Parameter | Value |
|---|---|
| N | 200 |
| M | 30 |
| E | 8 |
| $N_{ep}$ | 16 |
| $N_{sp}$ | 7 |
| $N_{gh}$ | 100 |

Table 3 The optimum PID controller coefficients according to BA with various objective functions

| Type of landing gear | $K_p$ | $K_i$ | $K_d$ |
|---|---|---|---|
| Nose landing gear/type 1 | 0.3568654 | 1.6379851 | 0.0253913 |
| Main landing gear/type 1 | 0.652985 | 1.753764 | 0.0469735 |
| Nose landing gear/type 2 | 0.8943572 | 4.63987612 | 0.08956345 |
| Main landing gear/type 2 | 0.1698754 | 1.2568432 | 0.06513287 |

**Calculation of the multi-objective function type 1** The multi-objective function for the displacements of body, nose, left and right gears is introduced by

$$[ITAE]_1 = \{ITAE\}_z + \{ITAE\}_\theta + \{ITAE\}_\varphi \\ + \{ITAE\}_{z_1} + \{ITAE\}_{z_2} + \{ITAE\}_{z_3} \quad (21)$$

The multi-objective function for the velocities of body, nose, left and right gears is given by

$$[ITAE]_2 = \{ITAE\}_{\dot{z}} + \{ITAE\}_{\dot{\theta}} + \{ITAE\}_{\dot{\varphi}} + \{ITAE\}_{\dot{z}_1} \quad (22) \\ + \{ITAE\}_{\dot{z}_2} + \{ITAE\}_{\dot{z}_3}$$

The multi-objective function for the accelerations of body, nose, left and right gears is given by

$$[ITAE]_3 = \{ITAE\}_{\ddot{z}} + \{ITAE\}_{\ddot{\theta}} + \{ITAE\}_{\ddot{\varphi}} \\ + \{ITAE\}_{\ddot{z}_1} + \{ITAE\}_{\ddot{z}_2} + \{ITAE\}_{\ddot{z}_3} \quad (23)$$

The total multi-objective function for the sum of displacements, velocities and accelerations is given by

$$[ITAE]_{Type1} = [ITAE]_1 + [ITAE]_2 + [ITAE]_3 \quad (24)$$

**Calculation of the multi-objective function type 2** The suspension impact force imparted to the fuselage from nose, left and right gears, respectively is given by

$$Fs_1 = ks_1[z-a\theta-(d-e)\varphi-z_1] + cs_1[\dot{z}-a\dot{\theta}-(d-e)\dot{\varphi}-\dot{z}_1] + f_1$$
$$Fs_2 = ks_2[z+b\theta-d\varphi-z_2] + cs_2[\dot{z}+b\dot{\theta}-d\dot{\varphi}-\dot{z}_2] + f_2$$
$$Fs_3 = ks_3[z+b\theta+e\varphi-z_3] + cs_3[\dot{z}+b\dot{\theta}+e\dot{\varphi}-\dot{z}_3] + f_3 \quad (25)$$

Table 4 The optimal suspension parameters according to BA with various objective functions

| Type of landing gear | $P_{sh}$ | $P_{sl}$ | $C_s$ | $K_s$ |
|---|---|---|---|---|
| Nose landing gear/type 1 | 15,946,257 | 98,653 | 87,952 | 1,235,469 |
| Main landing gear/type 1 | 17,945,366 | 99,136 | 91,325 | 1,495,631 |
| Nose landing gear/type 2 | 16,431,298 | 95,431 | 82,651 | 1,132,781 |
| Main landing gear/type 2 | 18,456,327 | 97,649 | 85,761 | 1,394,218 |

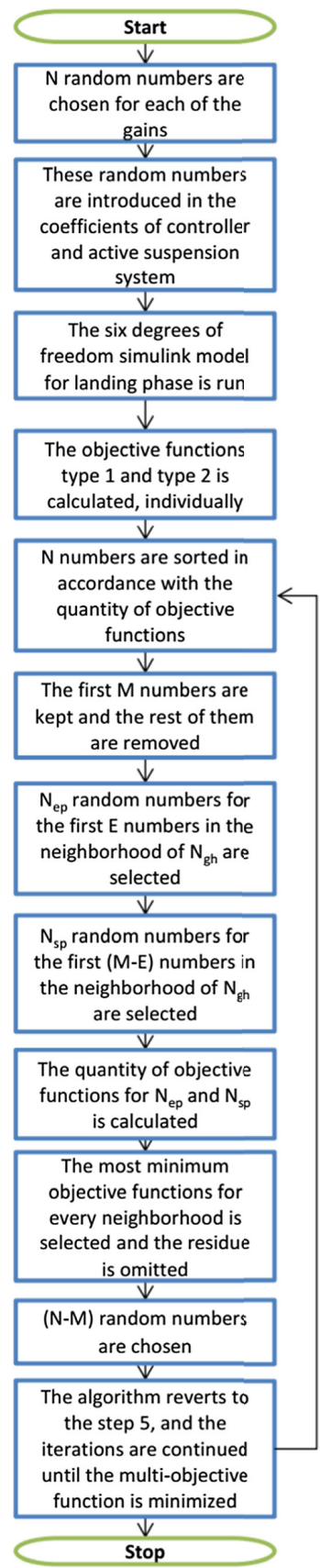

Fig. 4 Flowchart of multi-objective function type 1 and type 2

The tyre impact force imparted to the nose, left and right gears from ground, respectively are given by

$$Ft_1 = kt_1 z_1$$
$$Ft_2 = kt_2 z_2 \quad (26)$$
$$Ft_3 = kt_3 z_3$$

The multi-objective function for the suspension impact force is given by

$$[ITAE]_1 = \{ITAE\}_{Fs_1} + \{ITAE\}_{Fs_2} + \{ITAE\}_{Fs_3} \quad (27)$$

The multi-objective function for the tyre impact force is given by

$$[ITAE]_2 = \{ITAE\}_{Ft_1} + \{ITAE\}_{Ft_2} + \{ITAE\}_{Ft_3} \quad (28)$$

The total multi-objective function for the sum of suspension and tyre impact forces is given by

$$[ITAE]_{Type2} = [ITAE]_1 + [ITAE]_2 \quad (29)$$

Table 5 The parameters used in the numerical simulation for an Airbus320–200 airplane model

| Symbol | Value |
|---|---|
| M | 64,500 |
| $m_1$ | 300 |
| $m_2$ | 300 |
| $m_3$ | 300 |
| $ks_1$ | 15e5 |
| $ks_2$ | 15e5 |
| $ks_3$ | 15e5 |
| $cs_1$ | 1e5 |
| $cs_2$ | 1e5 |
| $cs_3$ | 1e5 |
| $kt_1$ | 3e6 |
| $kt_2$ | 3e6 |
| $kt_3$ | 3e6 |
| $I_{yy}$ | 1,278,370 |
| $I_{xx}$ | 3,781,268 |
| a | 10.88 |
| b | 1.76 |
| d | 3.795 |
| e | 3.795 |

## 4 Numerical simulation

In this section, an Airbus320–200 airplane vibrational model (type: conventional passenger) is studied in the control system simulation software as the case study. The values of model parameters have been used according to Table 5 (Doren 2009). The dynamic responses for this model by using the numerical simulation in MATLAB Simulink environment are obtained in the touchdown phase with three different scenarios of landing. In this paper, the sink speed is 3 m/s for this special aircraft.

### 4.1 Dynamic response for three points landing

In three points landing according to Fig. 6, landing gears hit with ground simultaneously and both of nose and main wheels absorb vertical energy communally. In this scenario, the aircraft lands with initial vertical velocity of 3 m/s (V) and without initial pitch and roll angles and the weight and lift forces are equal in magnitude (Currey 1998). In Fig. 6, L, W and T are the lift aerodynamic force, weight of the aircraft and

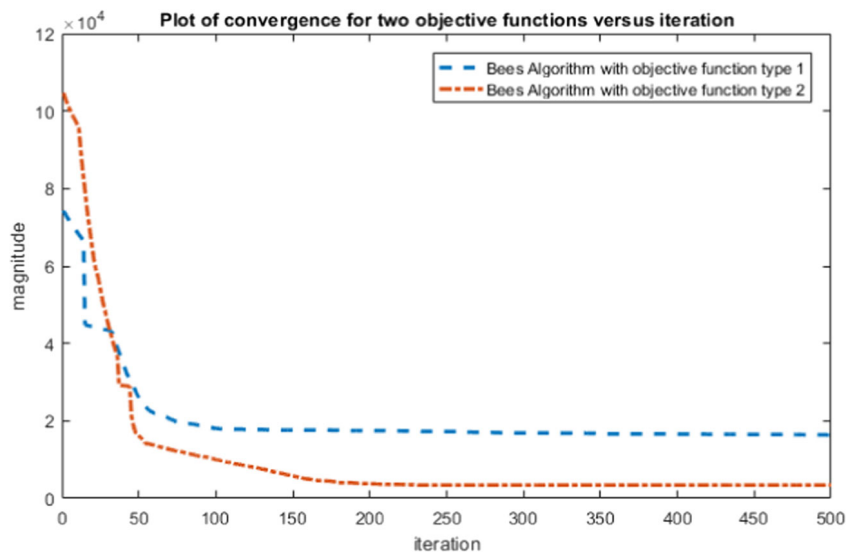

Fig. 5 Magnitude of two objective functions versus iteration

<em>Improved design of an active landing gear for a passenger aircraft using multi-objective optimization...</em>

**Fig. 6** Level load case for three point touchdown (Roskam 1989)

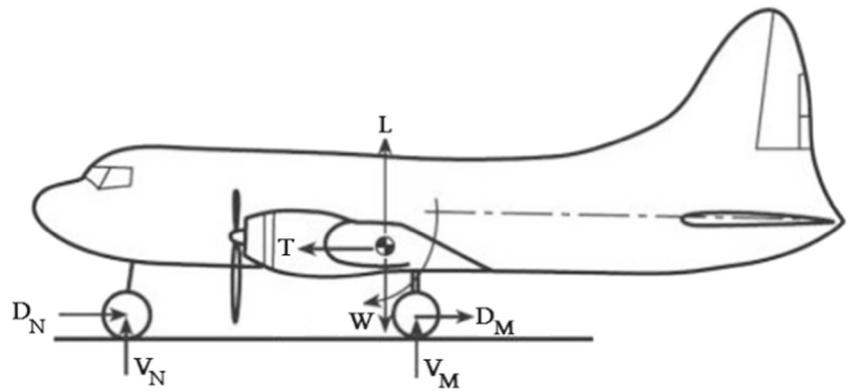

engine force, respectively. $D_N$ and $D_M$ represent the drag aerodynamic forces. $V_N$ and $V_M$ introduce the vertical forces.

Here, the input variable is touchdown speed and the output variables are the bounce displacement, bounce momentum, pitch displacement, pitch momentum, suspension travel and suspension impact force. These output variables are shown in Figs. 7, 8, 9, 10, 11 and 12 for time-domain and frequency-domain using signal processing by Fast Fourier Transform (El-Shafie et al. 2012). The comparison is performed between four suspension subsystem performances consist of passive, active based on Ziegler-Nichols, active on the basis of bee algorithm with two objective functions type 1 and type 2 as stated previous section.

All the simulation results demonstrate that, the Root Mean Square (RMS) values of these important parameters as the best index for comparison of dynamic behavior (Bissell and Chapman 1992) have been considerably decreased which is due to the optimal dynamic active control unit added to optimum suspension system. As well as, frequency-domain responses using signal processing by Fast Fourier Transform as Power Spectral Density (PSD) of outputs prove and validate the higher comfort level for passengers. According to Tables 6 and 7 and with attention to improvement percentage, the superior performance for all of dynamic parameters is obtained by using Bees Algorithm on the basis of objective function type 2.

### 4.2 Dynamic response for the sensitivity analysis of three points landing

From Figs. 13, 14, 15, 16, 17 and 18 simulation results as bounce displacement and momentum, pitch displacement and momentum, suspension travel and impact force in time-domain and frequency-domain as validation of performance improvement for dynamic responses according to bee algorithm on the basis of two multi-objective functions, it can be deduced that the reasonable estimating error of the sprung mass due to the additional payload and the touchdown speed because of environment condition when landing like wind have little effect on the performance of BA-based PID and the active suspension system efficiency changes less than 10% that shows low sensitivity (good performance) of this aircraft in this condition. The comparison of these results according to Tables 8 and 9 shows that the best performance for the active suspension system is on the basis of bee algorithm according to multi-objective function type 2.

**Fig. 7** Dynamic response of body displacement for three point landing

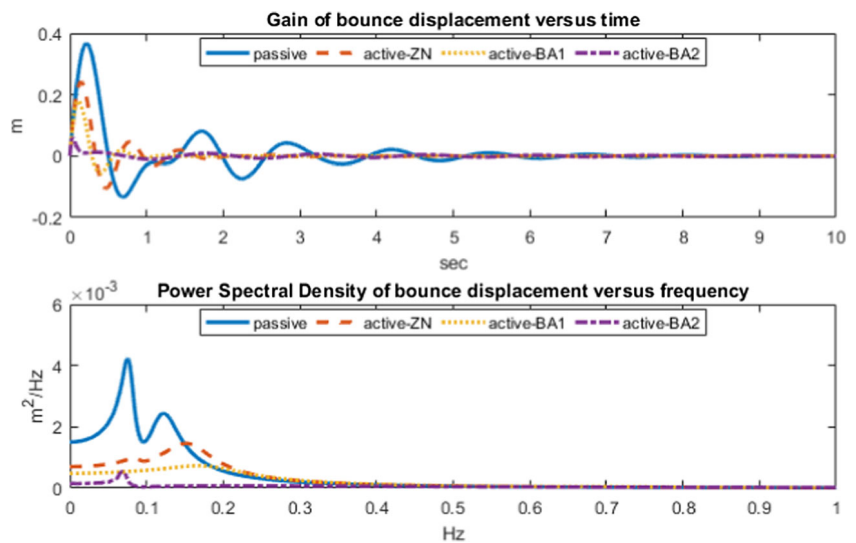

<em>Springer</em>

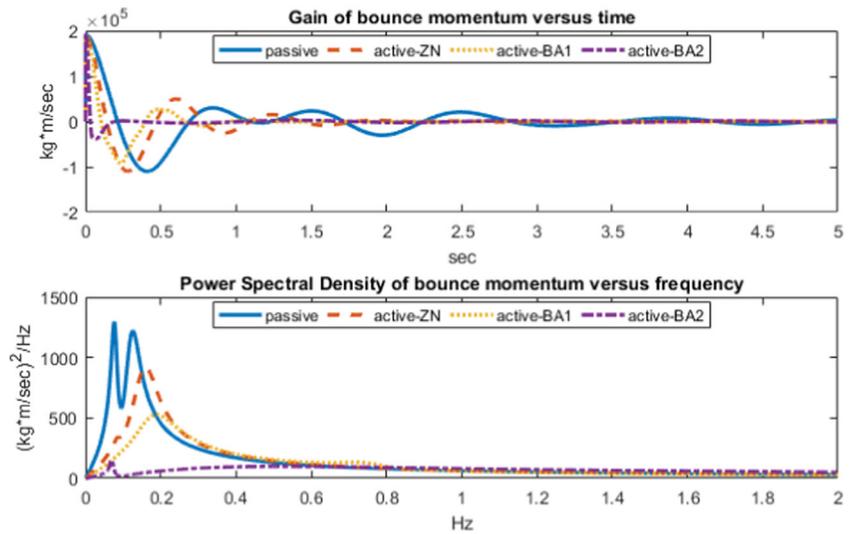

**Fig. 8** Dynamic response of body momentum for three point landing

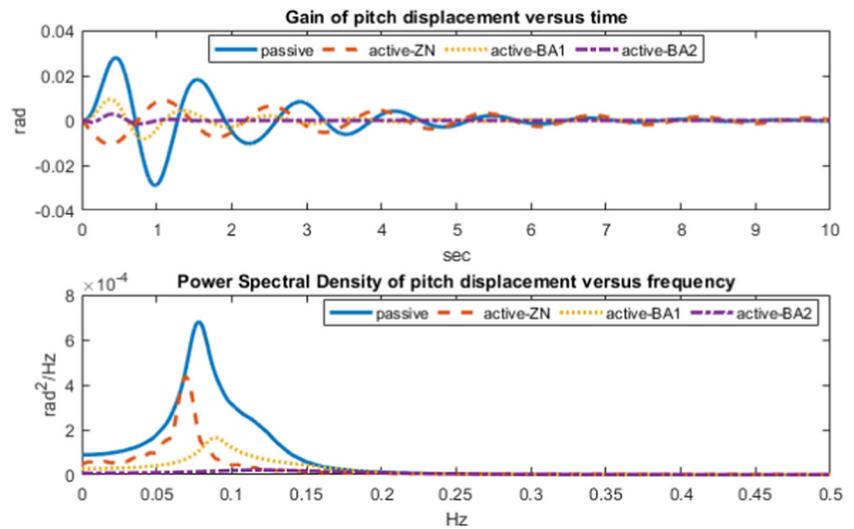

**Fig. 9** Dynamic response of pitch displacement for three point landing

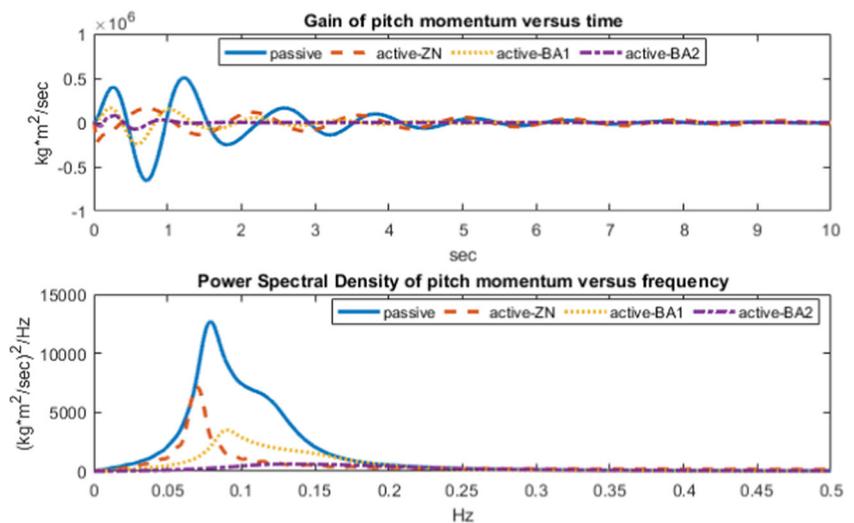

**Fig. 10** Dynamic response of pitch momentum for three point landing

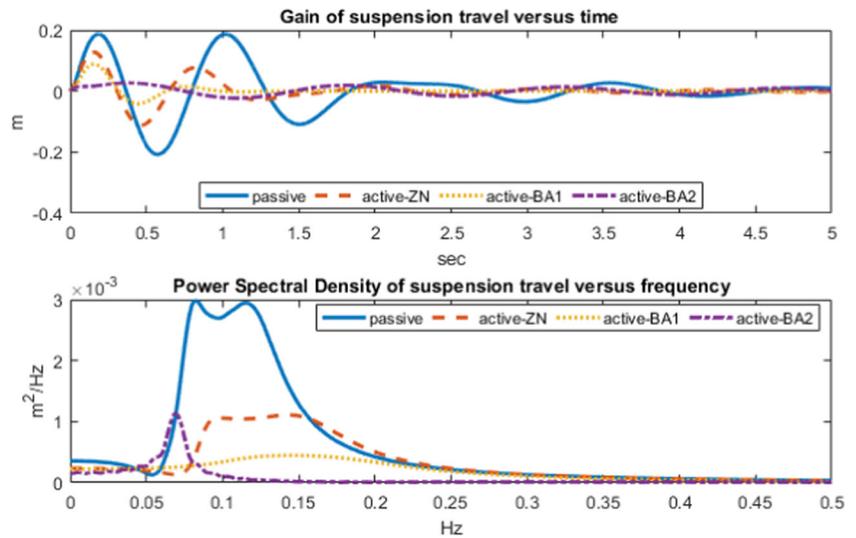

Fig. 11 Dynamic response of vibration absorber travel for three point landing

## 4.3 Dynamic response for the robustness analysis of two points landing

In two point landing according to Fig. 19, firstly, the main landing gears touch with the ground then nose landing gear touches with the ground so the majority of vertical energy at the touchdown moment absorbs by the main wheels. The most critical type of two point landing occurs when the aircraft pull downs horizontal tail completely. According to standard, the angular in this type landing is corresponding to the maximum of stall angle (Daniels 1996; Howe 2004). In this scenario, the aircraft lands with initial vertical velocity of 3 m/s (V) and the initial pitch angular of 12 degrees ($\alpha$) without initial roll angular with this consideration that the weight and lift forces are equal in magnitude. In Fig. 19, L and W are the lift aerodynamic force and the weight of the aircraft. $D_M$ represents the drag aerodynamic forces imparted to the main landing gears. $V_M$ introduces the vertical forces induced to the main landing gears.

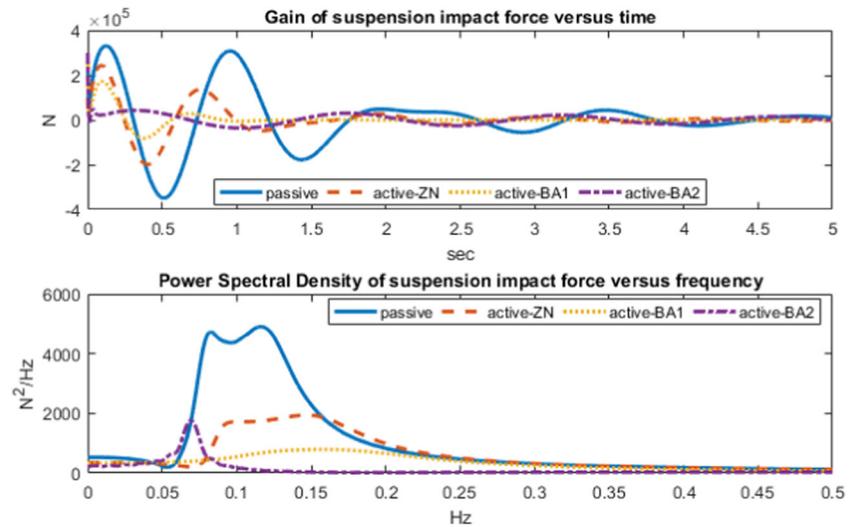

Fig. 12 Dynamic response of vibration absorber impact force for three point landing

Table 6 Comparison of RMS values for level load case

| Parameters | Performances | | | |
|---|---|---|---|---|
| | Passive | Active-ZN | Active-BA1 | Active-BA2 |
| Bounce displacement (m) | 0.0643 | 0.0344 | 0.0220 | 0.0062 |
| Bounce momentum (kg*m/s) | 2.6240e4 | 2.1692e4 | 1.6822e4 | 8.3978e3 |
| Pitch displacement (rad) | 0.0080 | 0.0037 | 0.0022 | 0.0005 |
| Pitch momentum (kg*m$^2$/s) | 1.5959e5 | 6.1059e4 | 5.0750e4 | 1.4606e4 |
| Suspension travel (m) | 0.0528 | 0.0245 | 0.0120 | 0.0096 |
| Suspension impact force (N) | 8.7877e4 | 4.3360e4 | 2.1936e4 | 1.5568e4 |

**Table 7** Comparison of improvement percentage for level load case

| Parameters | Performances | | | |
|---|---|---|---|---|
| | ZN to passive | BA1 to passive | BA2 to passive | Superior performance |
| Bounce displacement | 46 | 65 | 90 | BA2 |
| Bounce momentum | 17 | 35 | 68 | BA2 |
| Pitch displacement | 53 | 72 | 93 | BA2 |
| Pitch momentum | 61 | 68 | 90 | BA2 |
| Suspension travel | 53 | 77 | 82 | BA2 |
| Suspension impact force | 50 | 75 | 82 | BA2 |

Here, the input variables are the touchdown speed and initial pitch angular. The output variables are the bounce displacement and the bounce momentum. These output variables are shown in Figs. 20 and 21 for time-domain and frequency-domain using signal processing by Fast Fourier Transform. The comparison is performed between three suspension

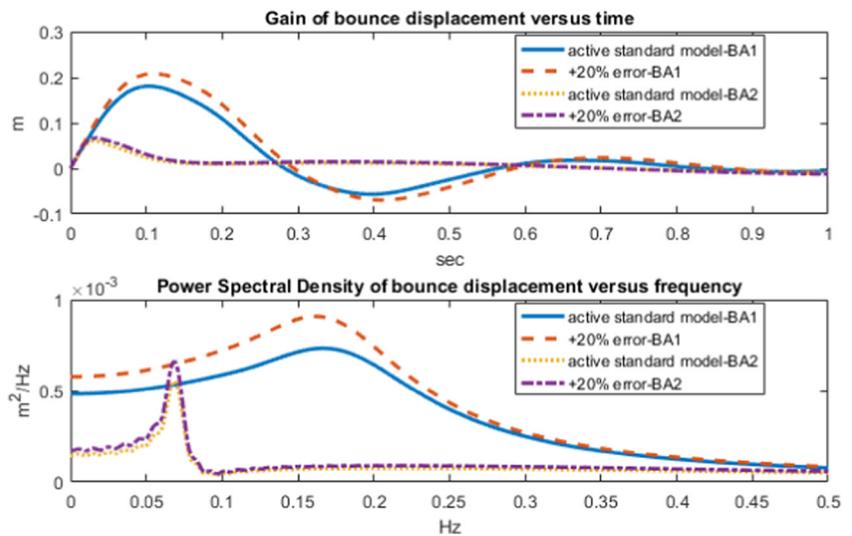

**Fig. 13** Dynamic response of body displacement for sensitivity investigation

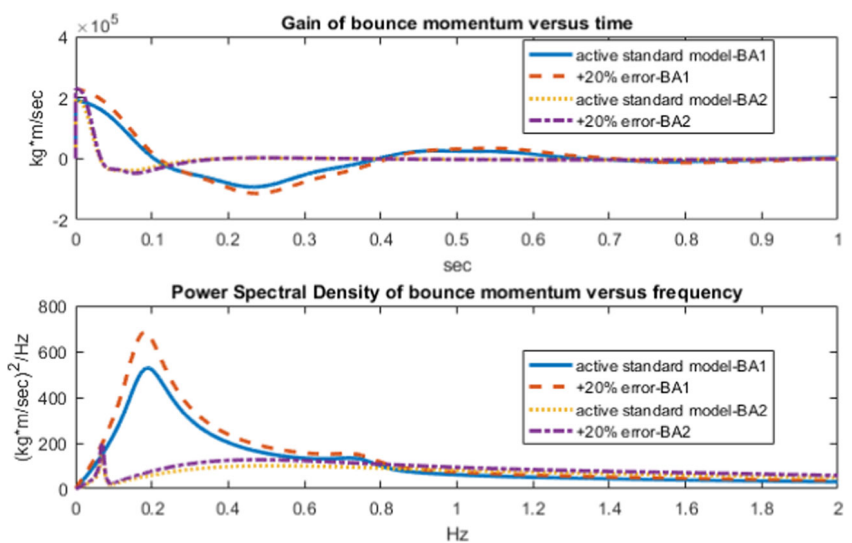

**Fig. 14** Dynamic response of body momentum for sensitivity investigation

Improved design of an active landing gear for a passenger aircraft using multi-objective optimization...

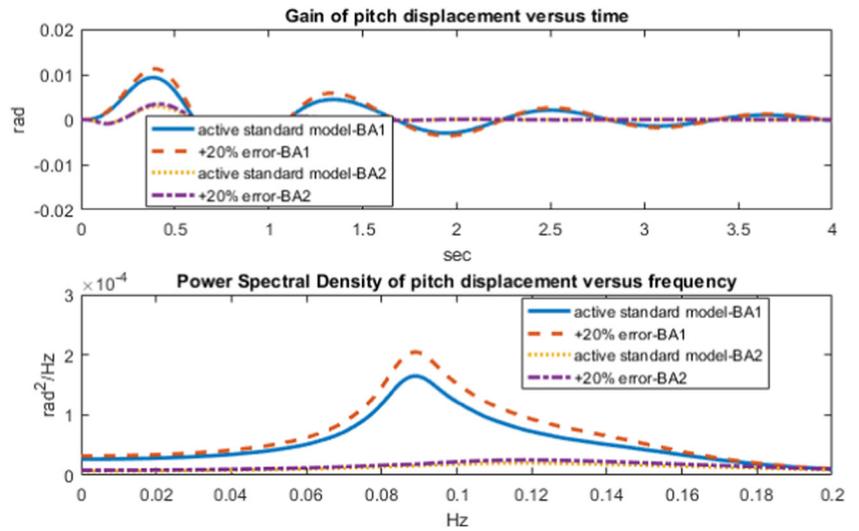

**Fig. 15** Dynamic response of pitch displacement for sensitivity investigation

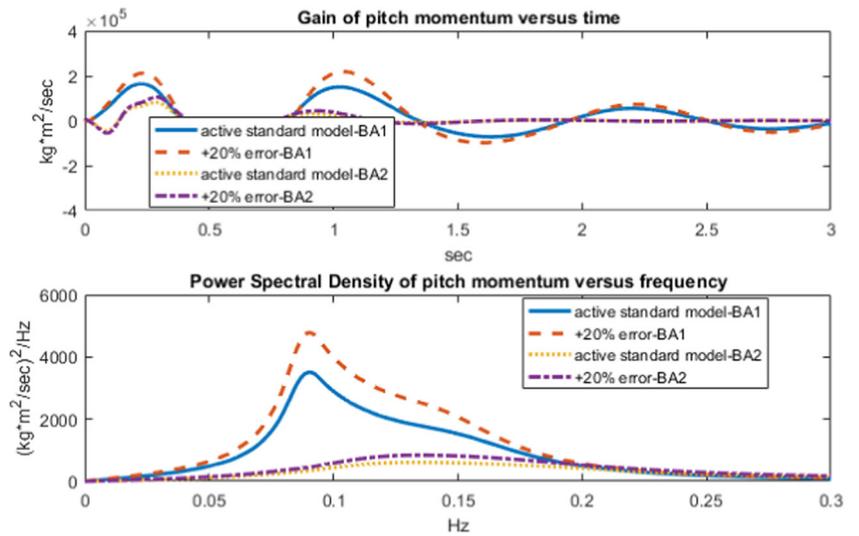

**Fig. 16** Dynamic response of pitch momentum for sensitivity investigation

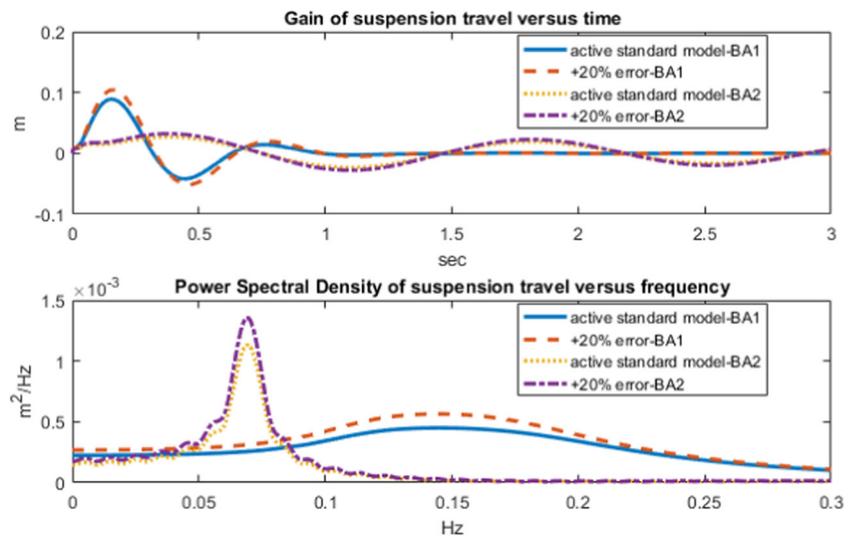

**Fig. 17** Dynamic response of vibration absorber travel for sensitivity investigation

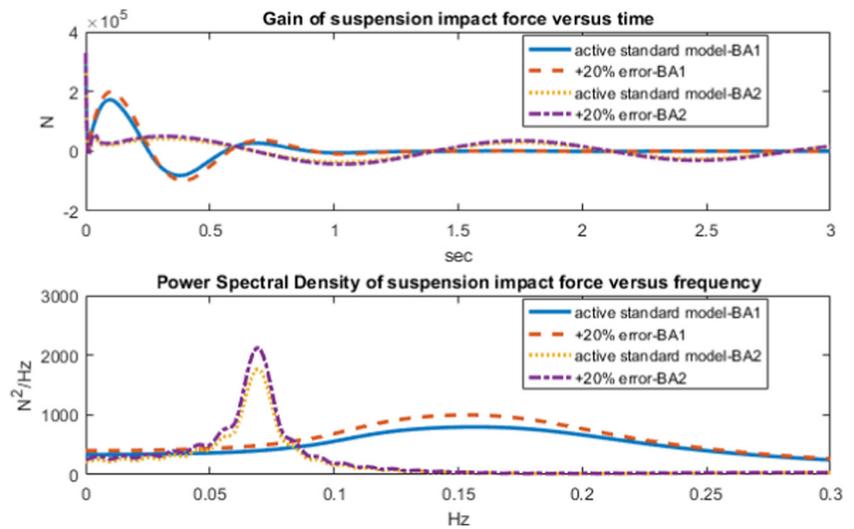

Fig. 18 Dynamic response of vibration absorber impact force for sensitivity investigation

subsystem performances consist of passive and active on the basis of bee algorithm with two objective functions type 1 and type 2.

The frequency-domain response using signal processing by Fast Fourier Transform demonstrates that the values of the power spectral density of bounce displacement and momentum are reduced in the whole frequency range for two types of multi-objective functions according to bee algorithm. According to Tables 10 and 11, for the aircraft body momentum and the bounce displacement, the values are reduced noticeably. It can improve the ride comfort importantly.

### 4.4 Dynamic response for the robustness analysis of one point landing

In one point landing according to Fig. 22, at the first, the left main landing gear touch with the ground then the right main landing gear does this and finally, the nose landing gear touches with the ground. The load of this type landing that can be caused by the whether inappropriate condition or landing on uneven runway is more than three points landing. The most critical case for one point touchdown take places when the airplane lands with maximum roll angle. The minimum magnitude of this angle in the conventional passenger aircraft is more or less 5 degrees (Catt et al. 1993). In this scenario, the aircraft lands with initial vertical velocity of 3 m/s (V) and initial pitch and roll angles of 12 and 5 degrees ($\alpha$, $\beta$), respectively (Daniels 1996; Howe 2004). The weight and lift forces are equal in magnitude. In Fig. 22, L and W are the lift aerodynamic force and weight of the aircraft.

Table 8 Comparison of RMS values for sensitivity investigation of level load case

| Parameters | Performances | | | |
|---|---|---|---|---|
| | Active-BA1 | +20% error-BA1 | Active-BA2 | +20% error-BA2 |
| Bounce displacement (m) | 0.0220 | 0.0263 | 0.0062 | 0.0074 |
| Bounce momentum (kg*m/s) | 1.6822e4 | 2.0818e4 | 8.3978e3 | 1.0127e4 |
| Pitch displacement (rad) | 0.0022 | 0.0027 | 0.0005 | 0.0006 |
| Pitch momentum (kg*m$^2$/s) | 5.0750e4 | 6.9239e4 | 1.4606e4 | 1.9347e4 |
| Suspension travel (m) | 0.0120 | 0.0145 | 0.0096 | 0.0116 |
| Suspension impact force (N) | 2.1936e4 | 2.6173e4 | 1.5568e4 | 1.8554e4 |

Table 9 Comparison of error percentage for sensitivity investigation of level load case

| Parameters | Performances | | |
|---|---|---|---|
| | Active-BA1 | Active-BA2 | Superior performance |
| Bounce displacement | 16.3 | 16.2 | BA2 |
| Bounce momentum | 19.2 | 17.1 | BA2 |
| Pitch displacement | 18.5 | 16.6 | BA2 |
| Pitch momentum | 26.7 | 24.5 | BA2 |
| Suspension travel | 17.2 | 17.2 | – |
| Suspension impact force | 16.2 | 16.1 | BA2 |

**Fig. 19** Tail down load case for two point touchdown (Roskam 1989)

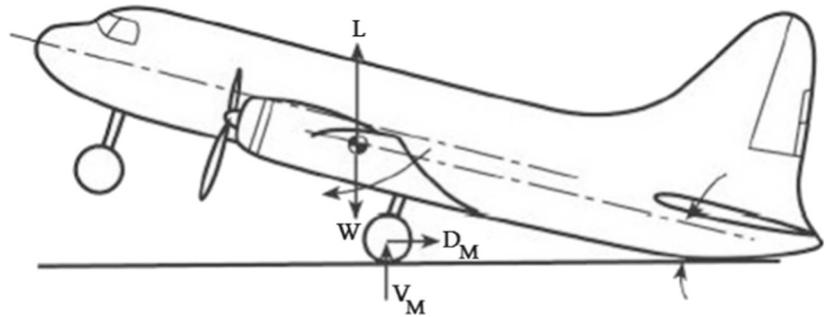

Here, the input variables are the touchdown speed and the initial pitch and roll angles. The output variables are the bounce displacement and the bounce momentum. These output variables are displayed in Figs. 23 and 24 for time-domain and frequency-domain using signal processing by Fast Fourier Transform. The comparison is performed between three suspension subsystem performances consist of the passive and active on the basis of bee algorithm with two objective functions type 1 and type 2.

In order to show the comparison of the Root Mean Square (RMS) of the body displacement and bounce momentum between the active suspension with two multi-objective functions on the basis of bee Algorithm and the passive performance, Table 12 gives the RMS of the various performances. From Table 13, two dynamic parameters for the active suspension are reduced compared with the customary performance with various multi-objective functions. The bounce displacement decreases by 74% at most, the Root Mean Square (RMS) of the bounce momentum decreases by 52% at most. As well as, the Figs. 23 and 24 give the Power Spectral Density (PSD) of the body displacement and the bounce momentum.

**Fig. 20** Dynamic response of body displacement for two point landing

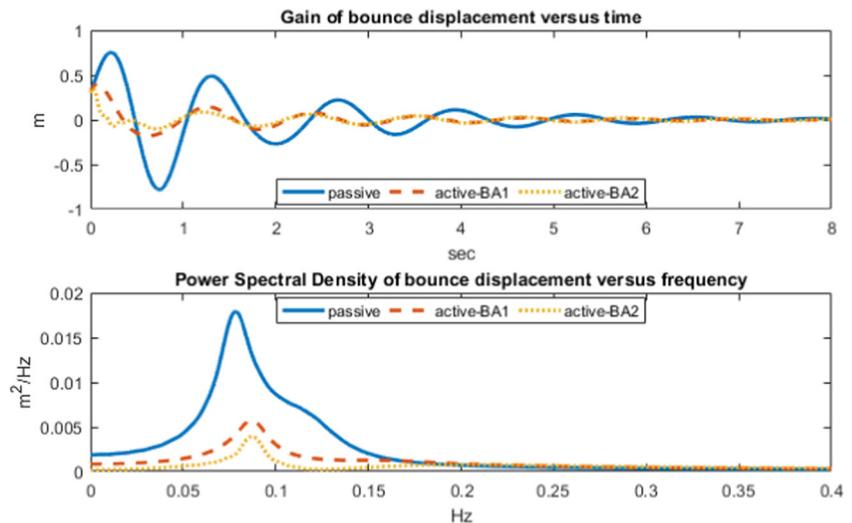

**Fig. 21** Dynamic response of body momentum for two point landing

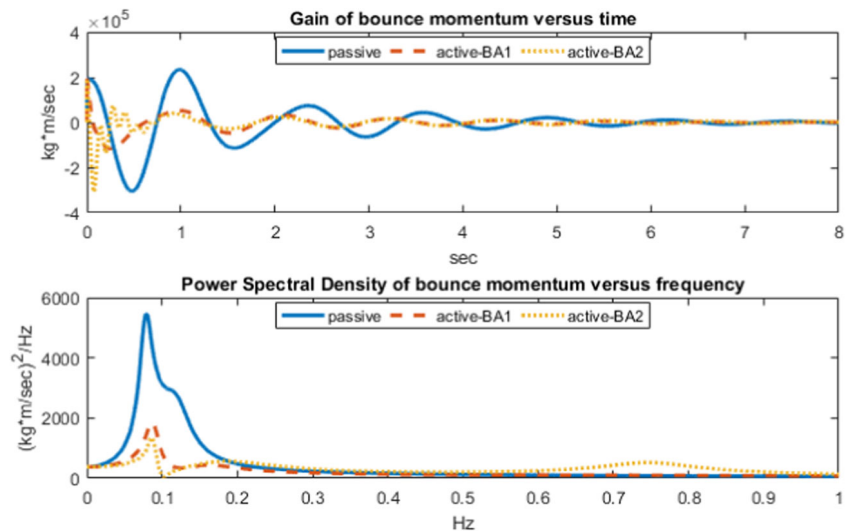

**Table 10** Comparison of RMS values for robustness investigation of tail down load case

| Parameters | Performances | | |
|---|---|---|---|
| | Passive | Active-BA1 | Active-BA2 |
| Bounce displacement (m) | 0.2128 | 0.0718 | 0.0430 |
| Bounce momentum (kg*m/s) | 7.1912e4 | 2.4720e4 | 2.7421e4 |

**Table 11** Comparison of improvement percentage for robustness investigation of tail down load case

| Parameters | Performances | | |
|---|---|---|---|
| | Active-BA1 to passive | Active-BA2 to passive | Superior performance |
| Bounce displacement | 66 | 80 | active-BA2 |
| Bounce momentum | 65 | 61 | active-BA1 |

## 5 Conclusion

The technical contribution of this paper has been the optimizing of the coefficients of controller, the parameters of hydraulic nonlinear actuator added to the traditional shock absorber system and the vibration absorber simultaneously by bee intelligent multi-objective algorithm. As well as, for proving the adaptability of this algorithm, this paper has presented the sensitivity analysis of three point landing due to the additional payload and touchdown speed and the robustness analysis of one and two point landings due to wind conditions as emergency situation on runway for an Airbus 320–200 aircraft as experimental case study.

This research has surveyed the optimal design of suspension system as one of the main subsystems and effective for the aircraft and the optimum structure of active control unit added on the shock absorber by using the Bee swarm-based optimization algorithm as the novelty of this research. This work achieves the consequence that, this optimized structure for landing gear and hydraulic nonlinear actuator for active performance of vibration absorber can significantly better the absorbing and damping process of the suspension system. Both simulation in time-domain and frequency-domain using signal processing by Fast Fourier Transform prove that the impact force during landing phase can be significantly curbed by the active suspension on the basis of bee algorithm according to two multi-objective functions in comparison with passive and Ziegler-Nichols method. The Root Mean Square (RMS) of the dynamic parameters can be diminished by 35% for level load case at least, the Root Mean Square of the dynamic responses can be diminished by 61% for tail down load case at least, and the Root Mean Square of dynamic parameters can be diminished by 38% at least for one wheel landing load case. Simultaneously, the values of the Power Spectral Density (PSD) of these dynamic parameters for level load case during landing phase, sensitivity investigation and robustness analysis for active vibration absorber for all performances mentioned are all reduced compared to the common passive shock absorber subsystem. The advantages of the bee algorithm are consisted of very efficient in finding optimal solutions and overcoming the problem of local optima.

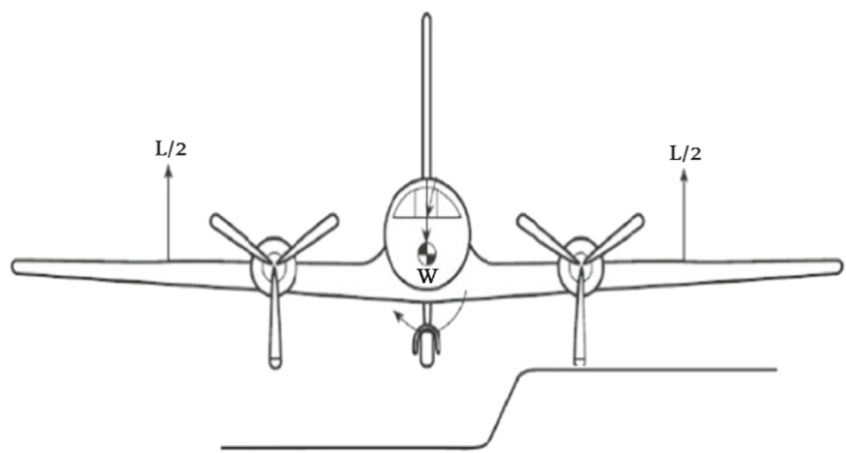

**Fig. 22** One wheel landing load case for one point touchdown (Roskam 1989)

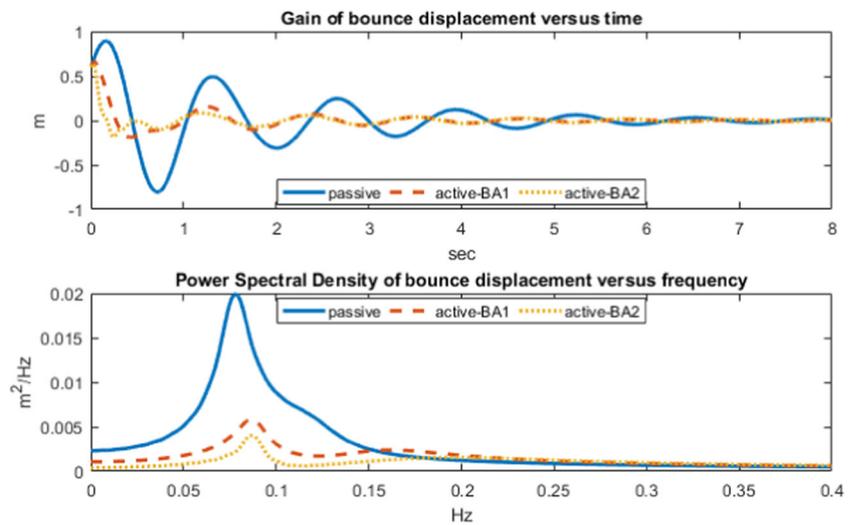

**Fig. 23** Dynamic response of body displacement for one point landing

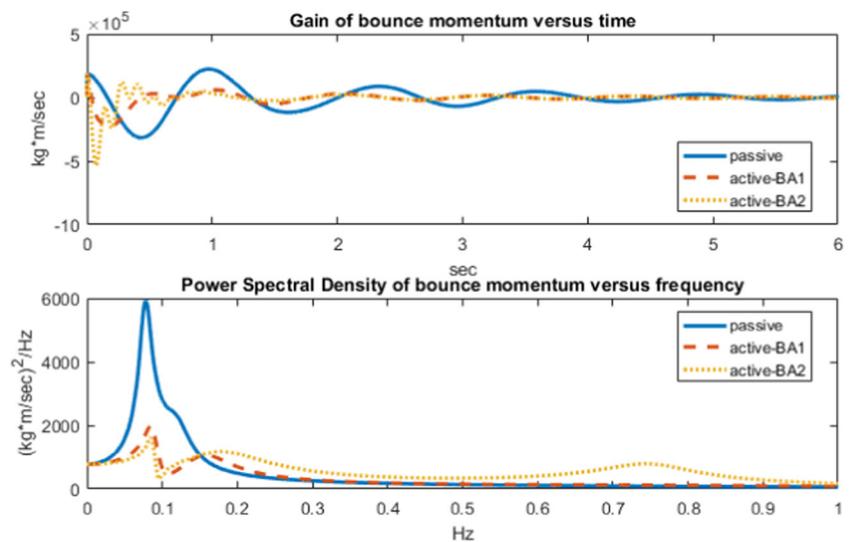

**Fig. 24** Dynamic response of body momentum for one point landing

## 6 Further works and new challenges

For the future research, other types of the conventional passenger aircrafts during touchdown phase under the scenarios of various landing or the taxiing phase under nonlinear and random runway excitation or combination of two phases will be studied. The smart swarm-based optimization algorithms will be implemented for robust, intelligent and nonlinear control techniques for the next work. As new challenges, these intelligent algorithms can be investigated for fault diagnosis in shock absorber system of aircraft using various neural networks. Remaining Useful Life (RUL) prediction allows for predictive maintenance of engineering systems, thus reducing costly unscheduled maintenance. Therefore, RUL prediction appears to be a hot issue attracting more and more attention as well as being of great challenge. Accurate remaining useful life prediction is critical to effective condition-based maintenance for improving reliability and reducing overall maintenance cost. Accurate health prognosis is critical for ensuring mechanical and aerospace systems reliability and reducing the overall life-cycle costs.

**Table 12** Comparison of RMS values for robustness investigation of one wheel landing load case

| Parameters | Performances | | |
|---|---|---|---|
| | Passive | Active-BA1 | Active-BA2 |
| Bounce displacement (m) | 0.2346 | 0.0894 | 0.0620 |
| Bounce momentum (kg*m/s) | 7.3763e4 | 3.5133e4 | 4.5592e4 |

**Table 13** Comparison of improvement percentage for robustness investigation of one wheel landing load case

| Parameters | Performances | | |
|---|---|---|---|
| | Active-BA1 to passive | Active-BA2 to passive | Superior performance |
| Bounce displacement | 62 | 74 | Active-BA2 |
| Bounce momentum | 52 | 38 | Active-BA1 |

## Compliance with ethical standards

**Conflict of interest**   The authors declare that they have no conflict of interest.

**Publisher's Note**   Springer Nature remains neutral with regard to jurisdictional claims in published maps and institutional affiliations.